\documentclass{elsart}

\usepackage{natbib}
\newcommand{\HI}{H{\,\small I}}
\newcommand{\FRI}{FR{-\small I}}
\newcommand{\FRII}{FR{-\small II}}

\usepackage{graphicx}

\usepackage{amssymb}

\begin{document}

\begin{frontmatter}



\title{The role of neutral hydrogen in radio galaxies}


\author[1,2]{Bjorn Emonts},
\author[2,3]{Raffaella Morganti},
\author[2,3]{Tom Oosterloo}

\address[1]{Department of Astronomy, Columbia University, 550 West 120th Street, New York, N.Y. 10027, USA}
\address[2]{Kapteyn Astronomical Institute, University of Groningen, P.O. Box 800, 9700 AV Groningen, the Netherlands}
\address[3]{Netherlands Foundation for Research in Astronomy, Postbus 2, 7990 AA Dwingeloo, the Netherlands}

\begin{abstract}
{We present morphological and statistical results of a study of neutral hydrogen (\HI) in a complete sample of nearby, non-cluster radio galaxies. We detect large-scale \HI\ emission in the early-type host galaxies of 25$\%$ of our sample sources. The large-scale \HI\ is mainly distributed in disk- and ring-like structures with sizes up to 190 kpc and masses up to $2 \times 10^{10} M_{\odot}$. All radio galaxies with $M_{\rm HI} \gtrsim 10^9 M_{\odot}$ have a compact radio source. When we compare our sample of radio-loud early-type galaxies with samples of radio-quiet early-type galaxies there appears to be no significant difference in \HI\ properties (mass, morphology and detection rate). This suggests that that the radio-loud phase could be just a short phase that occurs at some point during the life-time of many, or even all, early-type galaxies.}
\end{abstract}


\end{frontmatter}


\section{Introduction}
\label{sec:3_intro}

Large-scale \HI\ is detected in a growing number of early-type galaxies. In these proceedings numerous cases are presented by Oosterloo et al. and Serra et al., while many more cases are known from the literature \citep[e.g.][]{mor06b,gor97,sch97}. In the majority of the known cases, the \HI\ is distributed in regular rotating disk- or ring-like structures, that can reach far beyond the optical host galaxy and have \HI\ masses up to a few times the \HI\ mass of the Milky Way. However, this could be an observational bias, given that irregular structures are more frequently observed if one has the sensitivity to trace low mass \HI\ structures of only a few million solar masses \citep{mor06b}. Two good explanations for the origin of large-scale \HI\ structures in early-type galaxies are gas-rich galaxy mergers and the cold accretion of circum-galactic gas.\\ 
\vspace{-3mm}\\
In case of a major merger between gas-rich galaxies, part of the gas is transported to the central region of the merging system, where a sudden burst of star formation is triggered \citep{mih96}. Another part of the gas is expelled in large-scale tidal features of low surface-brightness, which can reach far beyond the optical host galaxy. If the environment is not too hostile and the gas in the tails remains gravitationally bound to the system, it can fall back onto the galaxy and settle into a disk- or ring-like structure within a few galactic orbits \citep[$>$1 Gyr;][]{bar02}. In the meanwhile, the stars in the merging systems have rearranged into an early-type galaxy \citep{hib96}.\\ 
\vspace{-3mm}\\
In case of cold accretion, galaxies accrete gas from the inter-galactic medium (IGM) via a cold mode, i.e. part of the gas cools along filamentary structures without being shock-heated \citep{ker05}. This gas (with $T < 10^5 K$) may cool further to form the large-scale structures of neutral hydrogen. On smaller scales, \citet{kau06} show that through the cooling of hot halo gas, cold gas can be assembled onto a galactic disc.

While both mechanisms provide a viable explanation for the formation of large-scale \HI\ around early-type galaxies, the exact formation mechanism is in most cases not evident from the \HI\ distribution alone. To verify the origin of these \HI\ structures it is therefore necessary to study other tracers of the formation history of the galaxy. A good tracer in this respect is the stellar population content of the galaxy. As described above, a major merger event triggers a burst of star formation in the host galaxy, which can be traced with optical spectra. We have used both \HI\ imaging and optical spectroscopy to study the formation history of nearby radio galaxies.

\section{\HI\ in radio galaxies}
\label{sec:3_sample}
\begin{figure}[!t]
\centering
\includegraphics[width=\textwidth]{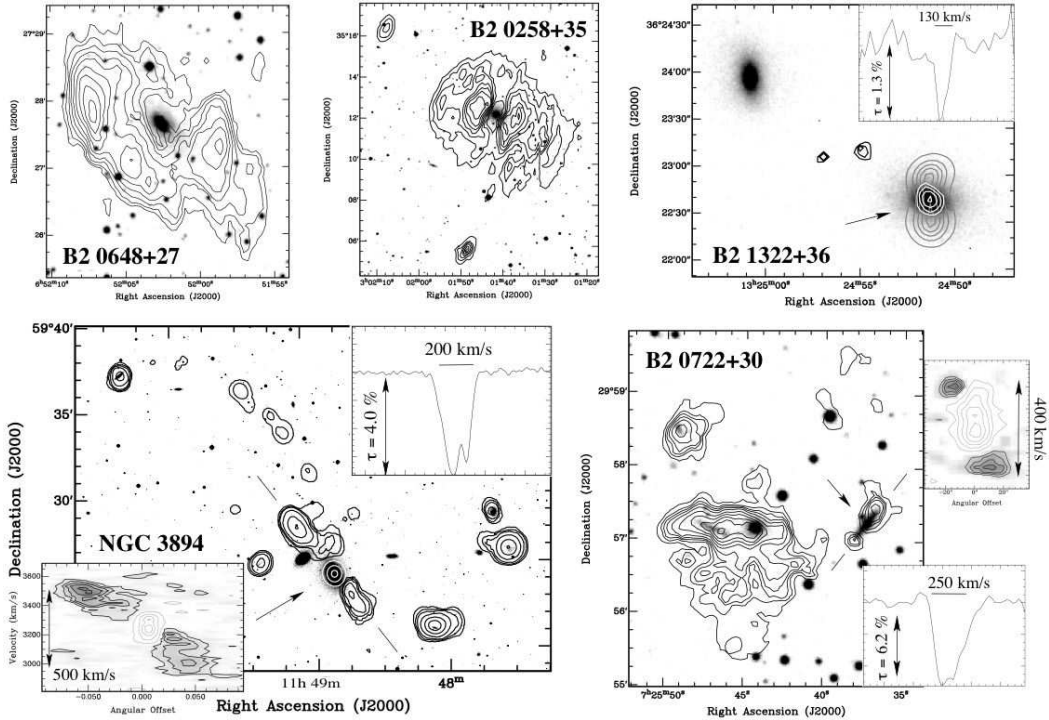}
\caption{From \citet{emo06b}. 0th-moment total intensity maps of the \HI\ emission (contours) in our \HI-rich nearby radio galaxies \citep[B2 1217+29, or NGC 4278, is presented by][]{mor06b}. Radio continuum is only shown for B2 1322+36 (grey contours); for the other sources, the radio continuum is unresolved (or only marginally resolved for B2 0722+30). Although \HI\ absorption is present in all five radio galaxies, we only show the \HI\ absorption (white contours/profile) in case it clarifies the morphology of the large-scale \HI. The arrows mark the host galaxies of our sample sources, while the broken lines show the direction along which the position-velocity (PV) plots are taken. {\sl Contour levels:} B2 0648+27: 0.22, 0.36, 0.52, 0.71, 0.95, 1.2, 1.5, 1.8, 2.1 $\times 10^{20}$ cm$^{-2}$ \citep[see also][]{emo06a}; B2 0258+35: from 0.34 to 3.0 in steps of 0.44 $\times 10^{20}$ cm$^{-2}$; B2 1322+36: 1.7, 2.3, 2.8 $\times 10^{20}$ cm$^{-2}$ (black) -- continuum: from 22 to 200 in steps of 44.5 mJy beam$^{-1}$ (grey); NGC 3894: 0.17, 0.49, 0.87, 1.7, 3.2, 4.6 $\times 10^{20}$ cm$^{-2}$ (black) -- PV: -1.0, -5.0, -10, -14 (grey), 1.0, 2.0, 3.0, 4.5, 6.5 (black) mJy beam$^{-1}$; B2 0722+30: 0.67, 1.3, 1.8, 2.3, 3.0, 4.0, 5.0, 6.0, 7.0, 8.0 $\times 10^{20}$ cm$^{-2}$ (part of the \HI\ disk that is observed in absorption is not plotted for clarification) -- PV plot: -0.5, -1.4, -2.4, -3.4, -4.4 (grey), 0.5, 0.7, 0.9, 1.1, 1.3 (black) mJy beam$^{-1}$.}
\label{fig:_sample}
\end{figure}
\begin{table}
\centering
\caption{{\sl \HI\ in radio galaxies.} Given is the name, NGC number, total \HI\ mass detected in emission, diameter of the \HI\ structure (or distance to the host galaxy for B2 1322+36), peak in \HI\ surface density, and morphology of the \HI\ structure (D = disk, R = ring, B = ``blob''). $H_{\circ} = 71$ km s$^{-1}$ Mpc$^{-1}$ used throughout this paper.}
\label{tab:hiradiogalaxies}
\begin{tabular}{llccccc}
$\#$ & B2 Name & NGC & M$_{\rm HI}$ & D$_{\rm HI}$ & $\Sigma_{\rm HI}$ & Mor. \\
 &  &  & (M$_{\odot}$) & (kpc) &(M$_{\odot}$/pc$^{2}$) & \HI \\
\hline
\hline
1 & 0258+35         & 1167 &  1.8$\times$10$^{10}$ & 160 & 2.7  & D\\ 
2 & 0648+27$^{\ a}$ & -     &  8.5$\times$10$^{9}$  & 190 & 1.7 & R \\
3 & 0722+30         & -     &  2.3$\times$10$^{8}$  & 15  & 4.1 & D \\
4 & 1217+29$^{\ b}$& 4278 &  6.9$\times$10$^{8}$  & 37 & - & D \\
5 & 1322+36         & 5141  &  6.9$\times$10$^{7}$  & 20  & 3.7 & B \\
6 & -               & 3894  &  2.2$\times$10$^{9}$  & 105 & 3.8 & R \\
\hline
\hline
\end{tabular}\\
\vspace{2mm} 
a). \citet{emo06a}; b). \citet{mor06b}.
\end{table}

Because major mergers are often invoked to trigger powerful radio sources \citep[e.g.][]{hec86}, it is particularly interesting to study the formation history of radio-loud early-type galaxies and compare this with that of radio-quiet early-type galaxies. For this reason we studied a complete sample of nearby radio galaxies in \HI, followed-up by an optical spectroscopic study of these systems (to study their stellar populations). In this paper we will focus on the \HI\ results and the comparison with \HI\ results on radio-quiet early-type galaxies. A more detailed analysis of the \HI\ properties of the individual radio galaxies is given in \citet{emo06a,emo06b}, while the stellar population analysis will be presented in a future paper.

Our \HI\ sample consists of 21 radio galaxies from the B2-catalogue ($F_{\rm 408MHz} \gtrsim 0.2$ Jy) up to a redshift of {\sl z} $\approx$ 0.04. This sample is {\sl complete}, with the restriction that we left out sources in dense cluster environments (since here large-scale gaseous features are likely wiped out on relatively short time scales) and BL-Lac objects. In addition we observed NGC 3894, which has a compact radio source with radio power comparable to our B2-sample sources. We leave NGC 3894 out of the statistical analysis in Sect. \ref{sec:3_radioquiet}. In total we observed 9 compact ($< 15$ kpc) radio sources and 13 extended ($> 15$ kpc) \citet{fan74} type-I radio sources. The sources have a radio power 22.0 $<$ Log ($P_{\rm 1.4\ GHz}$) $<$ 25.0 with no bias in $P_{\rm 1.4\ GHz}$ between the compact and extended sources. The radio sources are hosted by {\sl early-type galaxies (E and S0)}. Observations were made during various observing runs in the period Nov. 2002 - Feb. 2005 with the Very Large Array (VLA) in C-configuration and the Westerbork Synthesis Radio Telescope (WSRT). A full description of the sample and observing details will be presented in a future paper.

\section{Results on our radio-loud sample}
\label{sec:3_radioloud} 

We detect large-scale \HI\ emission in six of our sample galaxies. Images and properties of the large-scale \HI\ structures are shown in Fig. \ref{fig:_sample} and Table \ref{tab:hiradiogalaxies}. In most cases the \HI\ is distributed in a fairly regular rotating disk or ring (with diameter up to 190 kpc and mass up to $2 \times 10^{10} M_{\odot}$), although a varying degree of asymmetry is still visible in these structures. For one of these radio galaxies -- B2 0648+27 -- we already confirmed a merger origin through both the detection of a post-starburst stellar population, that dominates the light throughout the optical host galaxy \citep[see][]{emo06a}, and the fact that plume- or tail-like structures appear in deep optical imaging \citep{hei94}. The merger event in B2 0648+27 must have happened more than a Gyr ago, after which the \HI\ gas that was expelled during the merger had the time to fall back onto the host galaxy and settle in the regular rotating ring that we observe. In case a merger event is confirmed also for the other \HI-rich systems (with $M_{\rm HI} > 10^9 M_{\odot}$), than also for these systems the regular kinematics of the \HI\ gas suggest that the \HI\ structures are old. It is striking that we find {\sl no} clear cases of {\sl ongoing} mergers (in the form of tidal \HI-tails, -bridges or -plumes) associated with our sample sources. In fact, regardless of  the formation mechanism of these structures (be it major mergers or cold accretion), the large-scale structures are much older than the current period of radio-AGN activity.  

\begin{figure}[t]
\centering
\includegraphics[width=0.58\textwidth]{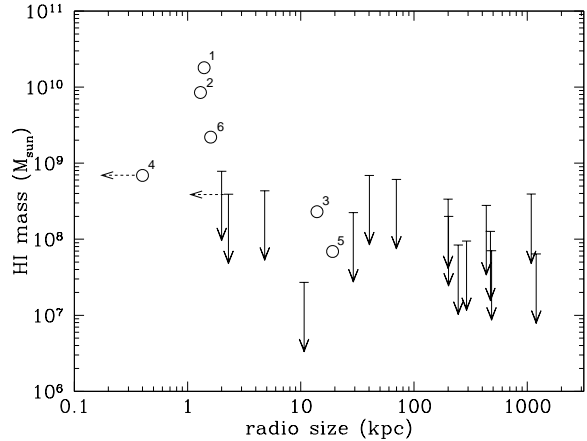}
\caption{Total \HI\ mass in emission plotted against the linear size of the radio sources. In case of non-detection a firm upper limit (3$\sigma$ across 400 km s$^{-1}$) is given.}
\label{HImasssize}
\end{figure}

Another interesting result is that we find a segregation in large-scale \HI\ mass content with radio source size (Fig. \ref{HImasssize}). The radio galaxies in our sample with $M_{\rm HI} \gtrsim 10^9 M_{\odot}$ all have a {\sl compact} radio source, while the more extended radio sources - all of Fanaroff $\&$ Riley type-I - do not contain these amounts of large-scale \HI. As explained in \citet{emo06b}, a possible explanation for this segregation is that - due to the re-distribution of the ISM in a merger event - the central radio sources in the \HI-rich radio galaxies do not grow, either because they are frustrated by ISM in the central region of the galaxy, or because the fuelling stops before the sources can expand. The lack of large amounts of \HI\ associated with  the extended \FRI\ sources suggests that they are likely fed through processes other than gas-rich mergers (e.g. cooling flows or the black hole's rotational energy). If confirmed by studies of larger samples, the neutral gas content may therefore be a specific property of the host galaxy for various types of radio sources.

\section{Comparison with radio-quiet samples}
\label{sec:3_radioquiet} 

\begin{figure}[b!]
\centering
\includegraphics[width=\textwidth]{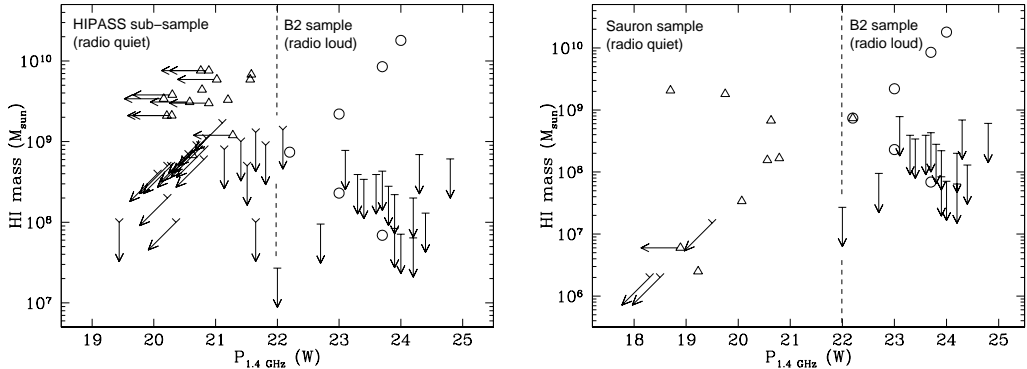}
\caption{\HI\ mass plotted against radio power for the early-type galaxies of the various samples. In case of non-detection the upper limit is plotted. The values of the HIPASS follow-up and the Sauron sample are taken from \citet{oos06} and \citet{mor06b}. For the B2 sample the circles represent the \HI\ detections and the flat arrows the non-detections; for the HIPASS follow-up and the Sauron sample the triangles represent the \HI\ detections and the pointed arrows the non-detections. The dividing line between the various samples is drawn for clarification and does not represent a physical division between radio-loud and radio-quiet galaxies.}
\label{fig:3_comparesamples}
\end{figure}

Recently, \citet{mor06b} and \citet{oos06} have completed two studies that were aimed at studying the occurrence and the morphology of large-scale \HI\ in early-type galaxies (not selected on radio loudness). In this Section we compare the results of these two studies with the results that we obtained on our sample of radio-loud early-type galaxies.\\ 
\vspace{-3mm}\\
{\sl HIPASS follow-up sample:} The first study by \citet{oos06} involves the follow-up imaging of \HI\ in early-type galaxies detected in the single-dish \HI\ Parkes All-Sky Survey (HIPASS). This project is described in detail in these proceedings by \citet{ser07}. The HIPASS sample of early-type galaxies is a complete sample with a typical detection limit of about $10^{9} M_{\odot}$. Initial results give a conservative \HI\ detection rate in early-type galaxies of $5 - 12 \%$ \citep{sad01}. Two-third of the \HI\ structures that are imaged in the HIPASS follow-up study are large and regular rotating disks or rings \citep{oos06,ser07}.\\ 
\vspace{-3mm}\\
{\sl Sauron sample:} The second study by \citet{mor06b} involves deep \HI\ imaging of 12 early-type galaxies selected from a larger, representative sample of early-type galaxies observed with the optical integral field spectrograph \textsc{SAURON}. With a low detection limit of a few $\times 10^{6} M_{\odot}$, the \HI\ detection rate in this sample is 70$\%$. The morphology of the \HI\ is more diverse than in the HIPASS follow-up study, with \HI\ morphologies ranging from regular rotating disks to irregular clouds, tails and complex distributions.

The \HI\ detection rate of our complete B2 sample of radio-loud early-type galaxies is 25$\%$. To compare this detection rate with the detection rates in the two samples of 'normal' early-type galaxies (i.e. not selected on radio-loudness), we plot in Fig. \ref{fig:3_comparesamples} the observed \HI\ mass  against the power of the radio source (in case of non-detection the upper limit is given). From Fig. \ref{fig:3_comparesamples} it is immediately clear that the early-type galaxies from the HIPASS and Sauron samples are radio-quiet compared with the radio galaxies in our B2 sample (one object common to both the B2 and the Sauron sample is the nearby radio galaxy B2 1217+29/NGC 4278). Table \ref{tab:detectionrates} summarizes the \HI\ detection rates of the three samples. Although the detection rates are very different for the three samples, we argue that this could be the result of a difference in sensitivity, rather than a true difference in \HI\ content. This is based on the fact that for the various samples there does not appear to be a significant difference in the percentage of galaxies with \HI\ masses above $10^{9} M_{\odot}$ (the detection limit of the HIPASS sample) and there is only a marginal difference for \HI\ masses above a few $\times 10^{8} M_{\odot}$ (the detection limit of the B2 sample) -- the latter effect beeing subject to the small number statistics of the Sauron sample.
\begin{table}
\centering
\caption{\HI\ detection rates of the various samples of early-type galaxies}
\label{tab:detectionrates}
\begin{tabular}{l|c|c|c}
 & HIPASS & B2 & Sauron \\
\hline
\hline
$\#$ galaxies & 818 & 20$^{*}$ & 12 \\
detection limit ($M_{\odot}$) & $\sim 10^{9}$ & ${\rm few} \times 10^{8}$ & ${\rm few} \times 10^{6}$ \\
detection rate ($\%$) & 5-12$^{**}$ & 25 & 70 \\
\hline
$\%$ with $M_{\rm HI} > 10^{9} M_{\odot}$ & 5-12 & 10 & 17 \\
$\%$ with $M_{\rm HI} > {\rm few} \times 10^{8} M_{\odot}$ & - & 25 & 33-50 \\
\hline
\hline
\end{tabular}\\
\vspace{2mm} 
\flushleft{{\small
$^{*}$ Complete B2 sample does not include NGC~3894 (see Sect. \ref{sec:3_sample}) and B2~1557+26 (which redshift of $z = 0.044$ is too high).\\
$^{**}$ Initial results for HIPASS, based on unconfused \HI\ detections \citep{sad01}.
}}
\end{table}

The morphology of the observed \HI\ structures in the two radio-quiet samples is remarkably similar to that of the \HI\ structures in our radio-loud B2 sample. In all samples, at the high-mass end ($M_{\rm HI} \gtrsim \times 10^{9} M_{\odot}$) the \HI\ is distributed in large and regular rotating disk- or ring-like structures. For lower amounts ($M_{\rm HI} \sim {\rm few} \times 10^{6} - 10^{8} M_{\odot}$), the samples also contain galaxies in which a more irregular \HI\ distribution is detected (as is the case for B2 1322+36).

Thus, as far as we can tell from the limited comparison between the three samples, there appears to be no major difference in both \HI\ detection rate and \HI\ morphology between the radio-quiet and radio-loud early-type galaxies in these samples. For sure, there is no evidence that our radio-loud sample has a higher detection-rate or contains more tidally distorted \HI\ structures than the radio-quiet samples. {\it If confirmed by larger samples with comparable sensitivity, this indicates that the radio-loud phase could be just a short period that occurs at some point during the lifetime of many -- or maybe even all? -- early-type galaxies.} 

As a final note, we would like to stress that our complete sample of radio-loud early-type galaxies did not include the more powerful radio sources of type \FRII, which are found at higher $z$ and which are often associated with major mergers \citep{hec86}.

\section{Conclusions}
\label{sec:conclusions} 

In a study of \HI\ in a complete sample of nearby, non-cluster radio galaxies, we detect large-scale \HI\ emission in 25$\%$ of the cases. The \HI\ is mainly distributed in fairly regular rotating disk- or ring-like structures. Regardless of the formation mechanism of these \HI\ structures (be it major mergers or cold accretion), their formation must have occurred long before the onset of the current phase of radio-AGN activity. We find no signs of ongoing mergers, nor do we find a major difference in morphology or detection rate with samples of radio-quiet early-type galaxies. If confirmed by larger samples, this indicates that the radio-loud phase could be just a short period that occurs at some point during the lifetime of many (all?) early-type galaxies.\\
\vspace{4.5mm}\\
{\bf Acknowledgements}\\
The author B. Emonts would like to thank Raffaella Morganti and Nanuschka Csonka for organising this nice workshop and Montse Villar-Mart\'{i}n for giving useful comments to improve the paper. Part of this project is funded by the Netherlands Organisation for Scientific Research (NWO) under Rubicon grant 680.50.0508.
\vspace{-5.4mm}




\end{document}